# Improving the understanding of the melting behaviour of Mo, Ta, and W at extreme pressures

Daniel Errandonea[1, *]

[1]*Departamento de Física Aplicada-ICMUV, Universitat de València,*
*Edificio de Investigación, c/Dr. Moliner 50, 46100 Burjassot (Valencia), Spain*

We discus existent conflicts between experimentally measured and theoretically calculated melting curves of Mo, Ta, and W. By assuming that vacancy formation plays a fundamental role in the melting process, an explanation for the measured melting curves is provided. Furthermore, we show that the Lindemann law fits well all the measured melting curves of bcc transition metals if the Grüneisen parameter is written as a power series of the interatomic distance. For completeness, we examine possible reasons for current disagreements between shock-wave and DAC experiments. To solve them, we propose the existence of an extra high P-T phase for Mo, Ta, and W.

Keywords: Transition metals, Melting at high pressure, Molybdenum, Tantalum, Tungsten

PACS numbers: 62.50.+p, 64.70.Dv, 71.20.Be, 61.10.Nz

---

[*] corresponding author, email: daniel.errandonea@uv.es, Te.: (34) 96 3543432, Fax: (34) 96 3543146

In materials science and geophysics, it is essential to understand how elements undergo the transition between solid and liquid phases at high pressures. For example, an accurate picture of materials performance under extreme conditions and detailed models of planetary interiors both depend on this knowledge. In recent years, the generation and measurement of simultaneous high pressures and high temperatures has undergone rapid development with the diamond-anvil cell (DAC) technique [1 – 3]. As a consequence of this, the amount of available data on melting behaviour at high pressure is increasing [4, 5]. The bcc transition metals (i.e. molybdenum (Mo), tantalum (Ta), and tungsten (W)) are especially interesting in this regard because they have very high melting points at ambient pressure and, in addition, at room temperature (RT) they remain in a stable body-centered cubic (bcc) structure up to extremely high pressures [6, 7]. Mo, Ta, and W therefore are valuable test cases for studying how materials melt. Despite a large number of experimental and theoretical efforts, however, agreement on the melting behaviour of Mo, Ta, and W at high pressure has been elusive. A number of experimental studies have been carried out with the use of shock-wave measurements [8 – 10] and laser-heated diamond-anvil cells [11, 12]. Likewise, theoretical studies have attempted to predict the melting curve of Mo, Ta, and W [13 - 17]. Unfortunately, substantial variation exists in the results of these earlier efforts owing to the use of different experimental and theoretical techniques.

The above-mentioned discrepancies are illustrated in Fig. 1 for the case of Ta. This figure presents the melting curves reported by different authors [11, 12, 13, 15, 18] including a new data point (3850 ± 150 K at 98 GPa) obtained by us using the technique described in Ref. [12]. It shows that at 30 GPa a melting temperature of 3600 ± 100 K was measured using the *in situ* speckle method [11] and synchrotron x-ray diffraction [12] whereas theoretical calculations [13] and Lindemann estimates [15] give a melting temperature ($T_m$) of 4500 K. In Fig. 1, it can be also seen that the disagreements even include the slope of the melting curve ($dT_m/dP$) at 1 bar (different estimates give a large dispersion of values, e.g. 24 K/GPa [12], 53 K/GPa [15], and 98 K/GPa [13]). On the other hand, ultra high-pressure shock experiments identified the melting of Ta in the pressure range from 2.5 Mbar (1 Mbar = 100 GPa) to 3 Mbar with $T_m$ estimated to be 7000 K – 1000K [9], while the extrapolation of the DAC data measured up to 1 Mbar gives $T_m$ ~ 4800 K in the same pressure range.

The disagreements resulting from different experimental and theoretical techniques are here debated for the cases of Mo, Ta, and W. Summing up, a phenomenological explanation for the melting curves measured using diamond-anvil cells is given, the Lindemann law is revisited, possible reasons for the overestimation of most of the theoretically calculated melting curves are examined, and the effect of superheating corrections on shock-melting data is discussed. Finally, for Mo, Ta, and W, the existence of an extra high P-T phase is proposed to solve discrepancies between shock-wave and DAC data.

We start our analysis with Mo, for which existent experimental and theoretical results have been summarized in Fig. 2. The melting curve of Mo has been measured using a laser-heated DAC up to near 1 Mbar, being found that Mo has a small melting slope at this pressure [11]. By comparison the Lindemann melting law [19],

$$\frac{\partial \ln T_M}{\partial \ln V_M} = \frac{2}{3} - 2\boldsymbol{g} , \qquad (1)$$

where γ = 1.59 [20] is the Grüneisen and $V_M$ is the molar volume (V) of the solid at melting, predicts a much steeper melting curve (see Fig. 2), being the estimated $T_m$ ~ 6000 K at 1 Mbar while DAC experiments gives $T_m$ ~ 3100 K. On the other hand, in the framework of melting as a dislocation-mediated phase transition [14, 16] a melting temperature close to 4200 K has been calculated at 1 Mbar (see Fig. 2) and *ab initio* molecular dynamics simulations found $T_m > 4000$ K at the same pressure [14].

In the case of the Lindemann law, the observed discrepancies are not surprising since it is an empirical law based on earlier experimental investigations of simple gases at low pressures. Clearly, noble gases characterized by a full valence shell of electrons ($s^2p^6$), which gives those elements special stability, are very different from the transition metals in which *d*-electrons play a dominant role.

In the case, of the dislocation-mediated melting model [14], an extrapolation of the bulk modulus to high pressure using only its first pressure derivative at ambient condition is assumed. The shear modulus is also similarly extrapolated to high pressures. Additional reasonable assumptions are considered by Burakovsky *et al.* to calculate the melting curve of Mo using their melting relation [14, 16], which has an accuracy of 17% at ambient pressure. Therefore, it is not unexpected that their model could lead to an overestimation of the melting temperature at high pressures. On top of this, it has been observed recently that the choose of different sets of ambient pressure parameters could lead to quite different results for the melting curve when using the dislocation-mediated melting model [21]. As a matter of fact, using this model Burakovsky *et al*. predicted for Mo a melting temperature of 4200 K at 1 Mbar, but Verma *et al.* found that $T_m > 5200$ K [22]. The differences between both calculations at higher pressures becomes comparable with their respective differences with the existent experiments [23], which suggests that the dislocation-mediated melting model is not a reliable approach for estimating the melting curves of Mo.

Let discuss know possible reasons for the differences between the DAC experiments [11] and the *ab initio* calculations [16]. It has been established that these calculations can accurately determine the melting curve of nearly free electron metals, such as aluminium (Al) [24], where the interatomic forces remain relatively unchanged upon melting [17]. However, this is not the case of bcc transition metals, where the electronic structure is rearranged upon melting [17, 25]. This different behaviour of bcc transition metals is quite significant for melting [17] and may be a possible cause for the large melting temperatures obtained from *ab initio* calculations, since these calculations does not take into account the free energy changes produced by the alterations of the *d*-electron band of Mo upon melting.

In contrast with others theoretical developments, it has been shown, that the flat melting curve observed in bcc transition metals can be understood with the help of a phenomenological vacancy-generation model of melting [12, 26, 27]. Within this framework, we calculated the melting curve of Mo using the Clausius-Clapeyron equation [28]:

$$\frac{\partial \ln T_M}{\partial P} = \frac{\Delta V_M}{\Delta H_M}, \qquad (2)$$

where $\Delta V_M$ and $\Delta H_M$ are, respectively, the difference in molar volume and enthalpy of the solid and liquid coexistent phases at melting conditions. Equation (2) was integrated following Ref. [12], assuming that the pressure dependence of $\Delta H_M$ is proportional to that of Ta and taken the ambient pressure values of $\Delta V_M$ and $\Delta H_M$ from the literature ($\Delta H_M = 40.3$ KJ mol$^{-1}$ and $\Delta V_M = 0.3$ cm$^3$ mol$^{-1}$) [29]. The obtained results are shown in Fig. 2, where it can be seen that the present model reproduces fairly well the experimental trend of the melting curve supporting the correctness of these data.

High-pressure melting data for Mo have been also obtained from shock experiments [8]. These experiments reported two transitions at 200 and 390 GPa; the first one was assigned to a solid-solid transition and the second one to melting. The calculated temperatures for these transitions were 3500 K and 10000 K, respectively. In contrast, extrapolation of static data [11] intersects the data point assigned to the solid-solid transition. In shock-wave experiments the occurrence of a structural transition is evidenced by a break in the shock-velocity particle-velocity plot. Thus, usually structural transitions are difficult to detect except for those transitions with very large volume changes, which is not the current case. This fact and the fact that in the shock-wave experiments reported in Ref. [8] the temperature was not measured but rather estimated (using some assumptions for the Grüneisen parameter and the specific heat) may be two of the reasons of the lack of agreement between shock-wave and DAC data. DAC measurements have also their own sources of uncertainties [30], but they are considerable smaller than in shock-wave experiments, which makes the DAC results more reliable. Another factor at play that may lead to an overestimation of the melting temperature in shock-wave experiments is the overshoot of the melting temperature due to the small time scale in shock experiments. Because of this, superheating ($\theta$) corrections must be applied to shock-wave data [31]. The shock-wave data plotted in Fig. 2 were obtained by correcting Hixson data [8] by considering $\theta = 0.3$. After the superheating correction, they are located at 3150 ± 500 K and 7700 ± 1500 K, in good agreement with recent estimates [16]. It can be seen in Fig. 2 that the superheating correction by itself cannot resolve the discrepancies between shock-data and DAC data. Even when considering that the estimated melting temperature was overestimated by 1000 K the lower bound of the temperature estimated for the shock-wave point at 390 GPa would be 5200 K, while the extrapolation of the present calculations and the DAC measurements [11] lays below 4000 K at the same pressure. Therefore, or the interpretation of the shock-data above summarized is in error or the P-T phase diagram of Mo has an unknown extra phase. The first scenario will imply that the first break in the shock-velocity particle-velocity plot would correspond to a melting transition from the solid to a very viscous liquid, maybe glass-like [32], being the second break due to a large viscosity change on the liquid. The second scenario will involve the existence of a triple point at the P-T conditions where the solid-solid boundary line detected in the shock experiments intercepts the melting curve (see Fig. 2). Such a triple point would produce a discontinuous change of the melting slope [4, 5], which could make converge the data measured below 1 Mbar [11] and the present calculations with the shock-wave data [8].

As already discussed above, in the case of Ta, both the Lindemann law ($\gamma$ = 1.69) and the theoretical calculations of Wang *et al.* [13] and Moriarty *et al.* [15] overestimate the melting curve. The existent theoretical and experimental data on Ta are shown in Fig. 1. There it can be seen the discrepancy between Lindemann, Wang, and Moriarty estimates with the DAC data [11, 12]. However, as in the case of Mo, the phenomenological model presented in Ref. [12] reproduces fairly well the measured melting curve. In this case, we recalculated the melting curve using more accurate values for the ambient pressure values of $\Delta V_M$ and $\Delta H_M$ ($\Delta H_M$ = 36 KJ mol$^{-1}$ and $\Delta V_M$ = 0.29 cm$^3$ mol$^{-1}$) [33]. The results obtained using these parameters are shown in Fig. 1, showing a good agreement with the experimental data.

In Fig. 1, the shock-wave melting point is also included [9] after the application of the superheating correction ($\theta$ = 0.3). As in the case of Mo, this data point has a temperature higher than the temperature extrapolated from the DAC data and the temperature calculated using the vacancy model at 300 GPa. However, the discrepancies are constrained to be smaller than 1000 K. The possible cause of this discrepancy will be discussed after discussing the case of W.

Tungsten is not the exception, and the differences between DAC experiments and theory observed in Mo and Ta are also observed on it (see Fig. 3). Fig. 3 summarizes the available experimental data [11, 34] together with the Lindemann estimates ($\gamma = 1.71$) and the dislocation-model calculations [14]. A new data point (4000 ± 100 K at 50 GPa) obtained using the technique described in Ref. [12] (open circle) is also included, confirming previous DAC experiments [11]. In the case of W, our phenomenological vacancy-model slightly overestimates the measured melting curve, but still reproduces the observed trend. The results of these calculations shown in Fig. 3 were obtained by extracting the ambient pressure values of $\Delta V_M$ and $\Delta H_M$ ($\Delta H_M = 35.4$ KJ mol$^{-1}$ and $\Delta V_M = 0.28$ cm$^3$ mol$^{-1}$) from Ref. [34].

The shock-wave melting data [10] obtained after applying the superheating correction is also shown in Fig. 3. In this case the difference between the shock-wave melting temperature and the temperature extrapolated from the static data at 400 GPa is larger then 2000 K. The fact that this discrepancy is systematically observed in Mo, Ta, and W (as well as in other transition metals like iron and vanadium) suggests that the existence of an unknown high P-T phase, proposed for the case of Mo, is the most likely origin of it. In the inset of Fig. 3 we present a possible schematic P-T phase diagram for these bcc transition metals. Validation of the proposed revised phase diagram can be obtained by the performance of new laser-heating x-ray diffraction measurements at megabar pressures.

When discussing differences between DAC melting data and theoretical calculations we pointed out the fact that the Lindemann expression, an empirical law based on low pressure data obtained from elements with a very different electronic structure than that of the transition metals, cannot be extrapolated to estimate the melting behaviour of transition metals at extreme pressures. The same can be said for *ab-initio* studies and other theoretical calculations. It does not seems especially meaningful to calculate the melting curve of the bcc transition metals at extreme pressures using models made for Al at 1 bar pressure just because they work satisfactory well in elements with a quite different electronic structure than the bcc transition metals.

To clarify the importance of the electronic structure issue on the melting behaviour under compression we show in Fig. 4 the melting curves of different elements along the periodic table reported by several authors. To facilitate the comparison among the several melting curves plotted, we represented the relative change of the melting temperature versus the relative change of the volume. All the data shown were taken from the literature [11, 12, 35 – 42] being the pressure-to-volume conversion made using well- known equation of states [37, 43 - 54]. In addition, the selected data correspond to a pressure range where melting is not affected by changes in the crystalline structure of the elements in order to simplify the discussion.

In Fig. 4, it can be seen that for all the studied bcc transition metals (i.e. Cr, V, Mo, Ta, and W) and those transition metals that having a crystalline structure different than bcc melt from bcc (i.e. Ti and Y) their melting curves follow a very similar trend. It is important to note that all these metals have a very similar electronic configuration (their *d* electronic shell is partially filled). On the contrary, the elements with a different electronic structure have a different behaviour, being possible to group them according their electronic configuration. In other words, the three noble gases (i.e. Ar, Kr, and Xe) with their characteristic closed shell configuration ($s^2p^6$) follow the same trend showing a very steep melting curve, nearly free electron *sp* metals like aluminium and magnesium have a similar behaviour, noble metals like copper and platinum show a quite similar melting curve, the same can be said the alkaline-earth metals calcium (Ca) and strontium (Sr), and for the magnetic transition metals nickel (Ni) and cobalt (Co). The systematic followed by Ni and Co is also followed by iron [11, 55].

In particular, it can be seen in Fig. 4 that not only the bcc transition metals, but also all the metals, which melt from the bcc structure, have a small melting slope. This is illustrated by the fact that Ca and Sr are the elements, which have the most similar behaviour to that of the bcc transition metals. In fact, these alkaline-earth metals remain also stable in the bcc phase in the compression range covered by Fig. 4 and under compression transform to early transition metals as a consequence of the pressure-driven *sp-d* electron transfer [56]. The small melting slopes shown by different elements which melts from bcc implies small volume changes at melting. This fact has been discussed before [11], being attributed to the fact that the bcc structure is more open than the close-packed structure. Basically, since the packing ratio for bcc (~0.68) is lower than for fcc or hcp (~0.74) it can be expected than the volume change for the melting of the bcc solid will be smaller than for a close-packed solid. Following this reasoning, a flattening of the melting curve should be observed for a solid with the simple-cubic structure (packing ratio ~0.52) , which is exactly what has been observed for Ca above 50 GPa [35].

As we mentioned above, we believed that many theoretical approaches fail to explain the experimentally observed behaviour for the melting curve of the bcc transition metals because there are based on models and approximations developed for elements with a complete different electronic configuration than that of bcc transition metals. This fact is supported by Fig. 4, which strongly suggests that the electronic configuration of an element is determinant on its melting behaviour under compression. Indeed, Ross *et al.* [17] have shown that omission of the *d*-band physics results in a large overestimation of the melting slope of bcc transition metals. These authors, proposed a semi-empirical model in which the *d*-band contribution to the total binding energy is described by the Friedel equation. The melting curve obtained using that model satisfactory agrees with the vacancy model estimates and the DAC measurements [11, 17]. On the other hand, in those calculations where melting is determined by the point where the Gibbs free energy of solid and liquid are equal, even a small error in the description of the energy difference can result in a large error in the melting temperature [30], being this fact a possible source for the observed discrepancies in the case of the bcc transition metals and a large number of materials as exemplified by the case of iron [55, 57].

Before closing this discussion, it is worth to mention that in Fig. 4 at a certain compression all the melting curves begin to flatten (i.e. they deviate from the linear behaviour predicted by the Lindemann law), this fact is in good agreement with the predictions of the phenomenological approach developed by Kechin [58] and seems to be a general feature of melting curves.

In the above discussion, we have shown that the Lindemann law tends to overestimate the melting temperature of bcc transition metals at high pressure. In the Lindemann estimates shown in Figs. 1 - 3 the Grüneisen parameter was assumed as pressure independent. However, the possibility of introducing a pressure dependence in $\gamma$ could reconcile the Lindemann law with the existent experimental data. In such a case the usual approximation consists in considering $\boldsymbol{g} = \boldsymbol{g}_0 \left(\frac{V}{V_0}\right)^n$, where $\gamma_0$ is the Grüneisen parameter at 1 bar and *n* is a constant parameter. However, this approximation does not work quite well. For Al, Cu, and Pt (metals with a steeper melting slope than Mo, Ta, and W), Wang *et al.* [51] after considering n values located within 0.5 – 1.5 have found that the approximation above described gives overestimated melting estimates, which are not compatible with experiments at any pressure. In the same way, it overestimates the melting curve for Mo, Ta, and W (see Fig 3). One possibility to solve the discrepancies between the Lindemann estimates and the DAC melting data is to consider a volume dependence of the Grüneisen parameter similar to that proposed by Burakovsky *et al.* [59]. The Grüneisen parameter can be modelled as:

$$\gamma(V) = \tfrac{1}{2} + g_1\left(\tfrac{V}{V_0}\right)^{1/3} + g_2\left(\tfrac{V}{V_0}\right)^q, \quad \gamma_1, \gamma_2, q = \text{const.}, \quad q > 1, \qquad (3)$$

through consideration of its low- and ultrahigh-pressure limits. This analytic form of γ was obtained under the assumption that (i) γ → ½ as V → 0, (ii) γ is an analytic function of the interatomic distance (i.e. $V^{1/3}$), and (iii) the first order coefficient of the Taylor-Maclaurin expansion of γ is not zero. The third term on the right-hand side of eq. (3) represents the quadratic and higher order terms in the power series. Thus, integrating eq. (1) it is straightforward to see that:

$$T_m(V) = T_{m0}\left(\tfrac{V}{V_0}\right)^{1/3} \exp\left\{-6g_1\left[\left(\tfrac{V}{V_0}\right)^{1/3} - 1\right] - \tfrac{2g_2}{q}\left[\left(\tfrac{V}{V_0}\right)^q - 1\right]\right\}. \qquad (4)$$

We used this expression to fit the experimental DAC data reported in Refs. [11], [12] and [40] and this work. Figs. 1 – 3 shows that by using the volume dependence of γ given by eq. (3) a fairly good agreement can be reached between the Lindemann law and the experimental data. The parameters obtained from the fits are given in Table I. The values obtained for γ, at 1 bar pressure, based on eq. (3) agree well with the literature data [20]. The values obtained for *q* are on the upper limit of the values obtained by Burakovsky [59] for several fcc metals, but they are compatible with them. In addition, the values obtained for γ at extreme pressures are quite reasonable. Particularly, for Mo at 3.9 Mbar we obtained γ ≈ 0.7, which is similar to the value estimated by Hixson (0.9) [8]. All these facts and the agreement obtained in the fits seems to indicate that eq. (4), deduced from the Lindemann law, might be a good analytical representation for the melting curve of bcc transition metals.

Summing up, we have discussed the present disagreements between different experimental and theoretical determinations of the melting curves of bcc transition metals. In particular, we showed that the melting curves of all transition metals melting from the bcc phase follow a clear systematic, which is quite different from that of elements with a different electronic configuration. Based on this fact, we presented possible reasons for the theoretical overestimation of their melting curves. Besides that, by considering a phenomenological vacancy-generation melting model we confirmed the melting curves determined in DAC measurements. This result suggests that theories of melting need to be further developed taking into account not only vacancies, but probably also point defects, dislocations, grain boundaries, and voids, which may play an important role in the bulk melting transition, as already shown for vanadium at 1 bar pressure [60]. The Lindemann law has been also discussed and a reformulation proposed to solve discrepancies between Lindemann estimates and experiments. Finally, the existence of a high P-T phase in Mo, Ta, and W has been proposed to make congruent the shock-wave and DAC melting data. We believe that more experimental and theoretical work is needed in order to resolve the discussed issues.

*Aknowledgments*

Daniel Errandonea acknowledges the financial support from the MCYT of Spain and the Universitat of València through the "Ramón y Cajal" program for young scientists. D.E. thanks all collaborators to his research on melting of metals, in particular the Mainz group (Max-Planck Institut für Chemie at Mainz), namely R. Boehler and the HPCAT (Advanced Photon Source at Argonne National Laboratory), namely D. Häusermann and M. Somayazulu. He also thanks M. Ross for stimulating discussions on melting of metals.

*References*

[1]  R. Boehler, Rev. Geophysics **38**, 221 (2000).

[2]  H. K. Mao and R. J. Hemley, Phil. Trans. R. Soc. Lond. A **354**, 1315 (1996).

[3]  G. Shen, M. L. Rivers, Y. Wang, and S. J. Sutton, Rev. Sci. Instrum. **72**, 1273 (2001).

[4]  R. Boehler, D. Errandonea, and M. Ross, *in High Pressure Phenomena*, eds. R. J. Hemley, G. L. Chiarotti, M. Bernasconi, and L. Ulivi (IOS Press, Amsterdam 2002), pg. 55.

[5]  R. Boehler, D. Errandonea, and M. Ross, High Pressure Research **22**, 479 (2002).

[6]  J. A. Moriarty, Phys. Rev. B **45**, 2004 (1992).

[7]  A. L. Ruoff, H. Xia, and Q. Xia, Rev. Sci. Instrum. **63**, 4342 (1992).

[8]  R. S. Hixson *et al.*, Phys. Rev. Letters **62**, 637 (1989).

[9]  J. M. Brown and J. W. Shaner, *in Shock Waves in Condensed Matter*, ed. By J. R. Asay, R. A. Graham, and G. K. Straub (Elsevier Science, New York 1983).

[10] R. S. Hixson and J. N. Fritz, J. Appl. Phys. **71**, 1271 (1992).

[11] D. Errandonea *et al.*, Phys. Rev. B **63**, 132104 (2001).

[12]  D. Errandonea, M. Somayazulu, D. Häusermann, and D. Mao, J. Phys.: Condens. Matter **15**, 7635 (2003).

[13] Y. Wang R. Ahuja, and B. Johansson, Phys. Rev. B **65**, 014104 (2002).

[14] L. Burakowsky, D. L. Preston, and R. R. Silbar, J. Appl. Phys. **88**, 6294 (2000).

[15] J. A. Moriarty, J. F. Belak, R. E. Rudd, P. Söderlind, F. H. Streitz, and L. H. Yang, J. Phys.: Condens. Matter **14**, 2825 (2002).

[16] A. B. Belonoshko, S. I. Sinak, A. E. Kochetov, B. Johansson, L. Burakovsky, and D. L. Preston, Phys. Rev. Lett. **92**, 195701 (2004).

[17] M. Ross, L. H. Yang, and R. Boehler, Phys. Rev. B **70**, 184112 (2004).

[18] N. S. Fatteva and L. F. Vereshchagin, Soviet Physics **16**, 322 (1971).

[19] F. A. Lindemann, Phys. Z **11**, 609 (1910).

[20] J. M. Wills and W. A. Harrison, Phys. Rev. B **28**, 4363 (1983).

[21] D. Errandonea, Journal of Physics: Condensed Matter **16**, 8801 (2004).

[22] A.K. Verma, R.S. Rao and B.K. Godwal, J. Phys.: Cond. Matter **16**, 4799 (2004).

[23] See Ref. [21] for a detailed comparison of the calculated melting curves reported in Ref. [16] and Ref.[22] and the experimental data reported in Ref. [11].

[24] L. Vocadlo and D. Alfe, Phys. Rev. B **65**, 214105 (2002).

[25] R. Wahrenberg, H. Stupp, H. G. Boyen, and p. Oelhafen, Europhys. Lett. **49**, 782 (2000).

[26] S. Mukherjee, R. E. Cohen, and O. Gülseren, J. Phys.: Condens. Matter **15**, 855 (2003).

[27] K. Kslazek and T Gorecki, High Temperatures – High Pressures **32**, 185 (2000).

[28] D. V. Schroeder, *An Introduction to Thermal Physics* (Addison Wesley, 1999).

[29] A. Fernadez-Guillermet, International Journal of Thermophysics **6**, 367 (1985).

[30] D. Alfe, L. Vocadlo, G. D. Price, and M. J. Gillan, J. Phys.: Cond. Matter **16**, S973 (2004).

[31] Sheng-Nian Luo and Th. J. Ahrens, J. Appl. Phys. **82**, 1836 (2003).

[32] V.V. Brazhkin *et al.*, Usp. Fiz. Nauk. **170**, 535 (2000).

[33] J. W. Shaner, G. R. Gathers, and C. Minichino, High Temperature High Pressures **9**, 331 (1977).

[34] A. Kloss, H. Hess, H. Schneidenbach, R. Grossjohann, Int. Journal of Thermophysics **20**, 1199 (1999).

[35] D. Errandonea, R. Boehler, and M. Ross, Phys. Rev. B **65**, 012108 (2002).

[36] R. Boehler and M. Ross, Earth Planet. Sci. Lett. **153**, 4589 (1997).

[37] R. Boehler, *in Recent Trends in High Pressure Research*, Ed. A. K. Singh (International Sciences, New York 1986).

[38] A. Kavner and R. Jeanloz, J. Appl. Phys. **83**, 7553 (1998)

[39] L. Vocadlo, D. Alfe, G. D. Price, and M. J. Gillan, J. Chem. Phys. **120**, 2872 (2004).

[40] D. Errandonea, R. Boehler, and M. Ross, Phys. Rev. Lett. 85, 3444 (2000).

[41] R. Boehler, M. Ross, P. Söderlind, and D. B. Boercker, Phys. Rev. Lett. **25**, 5731 (2001).

[42] P. Lazor, *Phase Diagrams, elasticity, and thermodynamics of Ni, Co, and Fe under high* pressure, Ph.D. Thesis, Uppsala University (1994).

[43] D. Errandonea, Y. Meng, D. Häusermann, and T. Uchida, J. Phys.: Condens. Matter **15**, 1277 (2003).

[44] H. Olijnyk and W. B. Holzapfel, Phys. Lett A **100**, 191 (1984).

[45] Y. Wang, D. Cheng, and X. Zhang, Phys. Rev. Lett. **84**, 3220 (2000).

[46] N. C. Holmes, J. A. Moriarty, G. R. Gathers, and W. J. Nellis, J. Appl. Phys. **66**, 2962 (1989).

[47] W. A. Grosshans and W. B. Holzapfel, Phys. Rev. B **45**, 5171 (1992).

[48] D. Errandonea, B. Schwager, R. Boehler, and M. Ross, Phys. Rev. B **65**, 214110 (2002).

[49] M. Ross *et al.*, J. Chem. Phys. **85**, 1028 (1986).

[50] C. S. Yoo, H. Cynn, P. Söderlind, and V. Iota, Phys. Rev. Lett. 84, 4132 (2000).

[51] Z. Wang, P. Lazor, and S. K. Saxena, Physica B **293**, 408 (2001).

[52] L. Ming and M. H. Manghnani, J. Appl. Phys. **49**, 208 (1978).

[53] A. Dewaele, P. Loubeyre, and M. Mezouar, Phys. Rev. B **69**, 092106 (2004).

[54] Y. Akahama, H. Kawamura, and T. L. Bihan, Phys. Rev. Lett. **87**, 275503 (2001).

[55] R. Boehler, Nature **363**, 534 (1993).

[56] H. L. Skriver, Phys. Rev. B **31**, 1909 (1983).

[57] D. Alfe, M. J. Gillan, and G. D. Price, Nature **401**, 462 (1999).

[58] V. V. Kechin, Phys. Rev. B **65**, 052102 (2001).

[59] L. Burakovsky, C. W. Greef, and D. L. Preston, Phys. Rev. B **67**, 094107 (2003).

[60] V. Sorkin, E. Polturak, and J. Adler, Phys. Rev. B **68**, 174102 (2003).

**Table I:** Best fit parameters of eq. (4) for different metals. The values taken from the literature for the Grüneisen parameter, the bulk modulus ($B_0$) and its pressure derivative ($B_0'$) are also given.

| Metal | $\gamma^a$ | $\gamma_0$ | $\gamma_1$ | $\gamma_2$ | q | $B_0$ [GPa] | $B_0'$ |
|---|---|---|---|---|---|---|---|
| W  | 1.71 | 1.62 | 3 10$^{-8}$   | 1.12 | 14 | 307[b] | 4.3[b]  |
| Ta | 1.69 | 1.62 | 1.2 10$^{-8}$ | 1.12 | 12 | 194[c] | 3.55[c] |
| Mo | 1.59 | 1.51 | 9.2 10$^{-9}$ | 1.01 | 13 | 267[b] | 4.4[b]  |
| V  | 1.27 | 1.25 | 2.4 10$^{-9}$ | 0.75 | 12 | 154[b] | 4.2[b]  |
| Cr | 1.24 | 1.23 | 4.9 10$^{-9}$ | 0.73 | 10 | 193[b] | 4.8[b]  |
| Ti | 1.1  | 1.08 | 4.8 10$^{-9}$ | 0.58 | 12 | 123[d] | 3.2[d]  |
| Y  | 1.05 | 1.03 | 2.3 10$^{-9}$ | 0.53 | 12 | 34[e]  | 5[e]    |

[a] Ref. [20], [b] Ref. [52], [c] Ref. [53], [d] Ref. [54], and [e] Ref. [47].

**Figure Captions**

**Figure 1:** Melting curve of Ta. DAC melting: (●) Ref. [11] and (o) Ref. [12] and new datum, shock melting (♦) [9] and piston-cylinder data (|) [18]. The dot-dashed lines correspond to theoretical calculations [13, 15] and the Lindemann estimates. The results obtained from the phenomenological melting model discussed in the text are illustrated by the solid line. The dotted line represents the estimates obtained from eq. (4).

**Figure 2:** Melting curve of Mo. DAC melting (●) Ref. [11]] and shock-wave data (♦) [8]. The dot-dashed lines correspond to theoretical calculations [16] and the Lindemann estimates. The results obtained from the phenomenological melting model discussed in the text are represented by the solid line. The dotted line represents the estimates obtained from eq. (4). The dashed lines represent proposed solid-solid and solid-liquid phase boundaries.

**Figure 3:** Melting curve of W. DAC melting (●) Ref. [11] and (o) new datum, shock melting (♦) [10], and wire-explosion melting (|) [34]. The dot-dashed lines correspond to theoretical calculations [14] and the Lindemann estimates. The results obtained from the phenomenological melting model discussed in the text are represented by the solid line. The dotted line represents the estimates obtained from eq. (4). The inset gives a schematic representation of the proposed phase diagram for the bcc transition metals.

**Figure 4:** Relative melting curves versus relative volume compression. The plotted data were extracted from the literature using reported melting curves [11, 12, 35 – 42] and well-known EOS [39, 43 - 54].

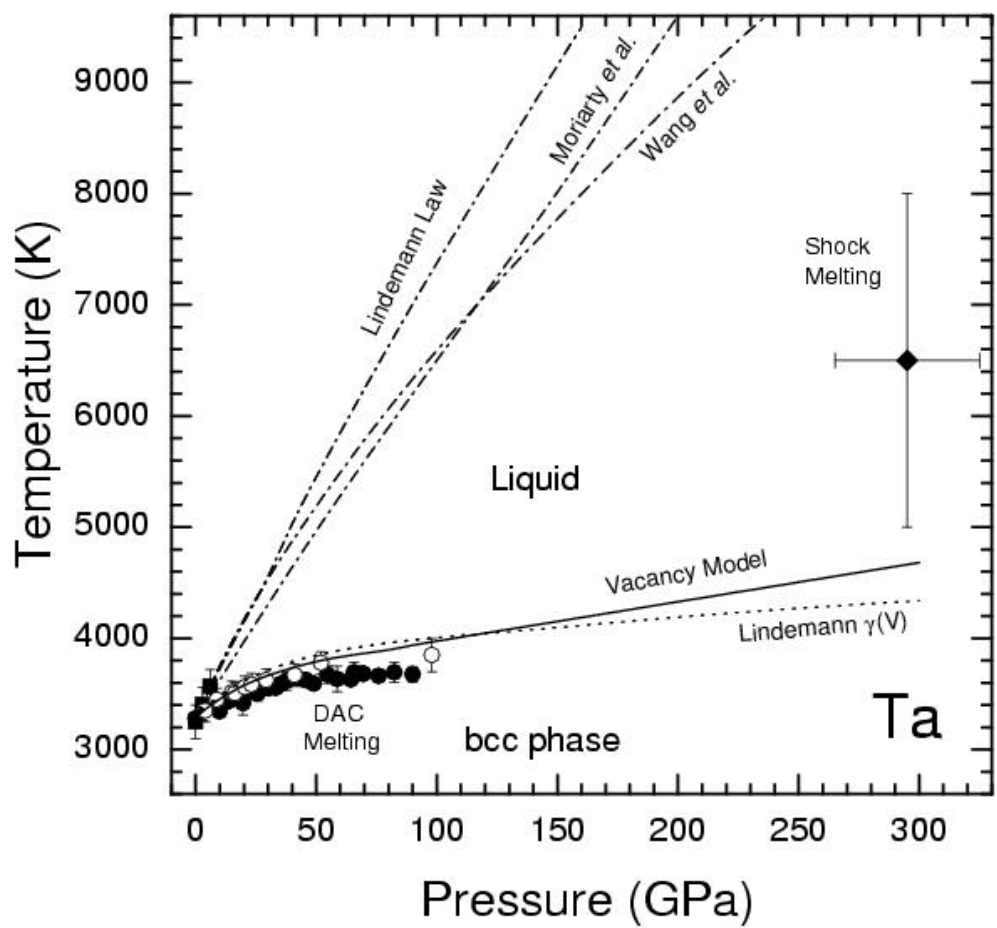

**Figure 1**

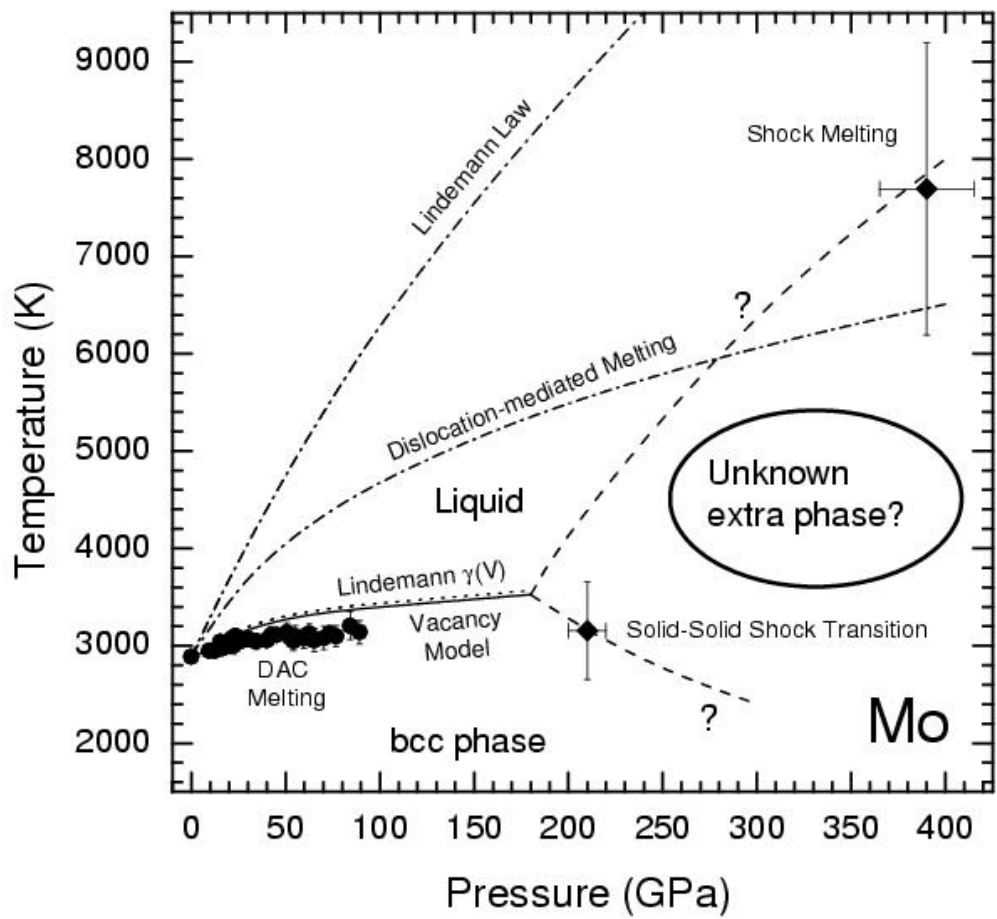

**Figure 2**

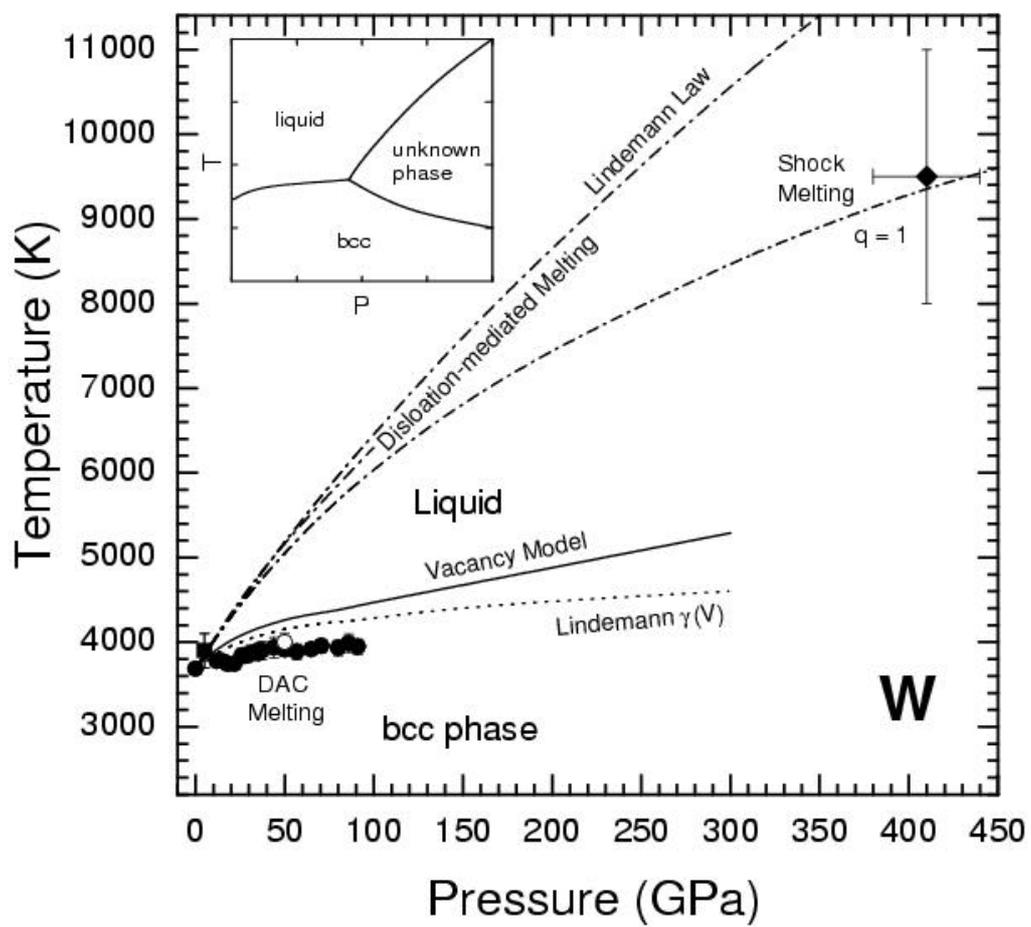

**Figure 3**

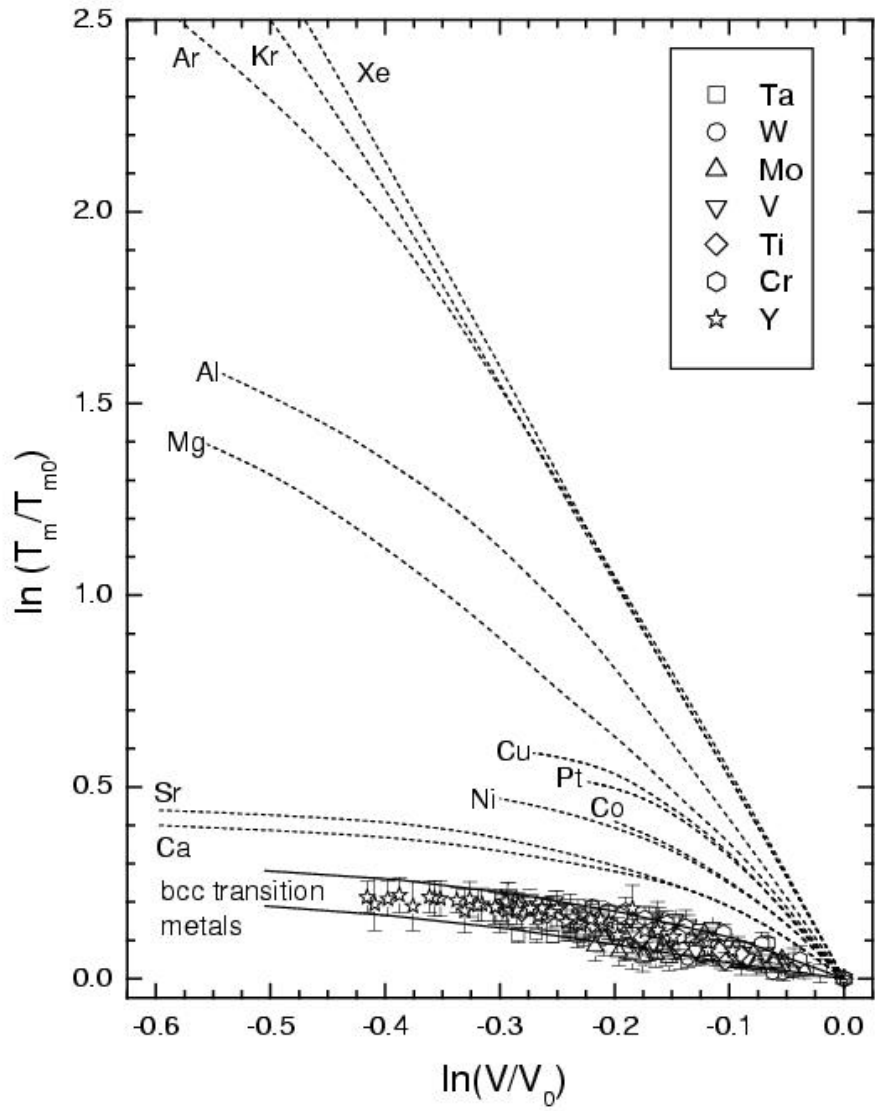

**Figure 4**